\documentclass[aip, jmp, amsmath,amssymb, preprint]{revtex4-2}

\usepackage{graphicx}
\usepackage{bm}
\usepackage{color}
\usepackage[colorlinks,linkcolor=blue,urlcolor=blue,citecolor=blue,bookmarks=false]{hyperref}

\newcommand{\argmax}{\mathop{\rm arg\,max}\limits}

\begin{document}

\title{Removing background and estimating a unit height of atomic steps from a scanning probe microscopy image using a statistical model}

\author{Yuhki Kohsaka}
\affiliation{RIKEN Center for Emergent Matter Science, Wako, Saitama 351-0198, Japan}
\email{kohsaka@riken.jp}
\date{\today}

\begin{abstract}
We present a statistical method to remove background and estimate a unit height of atomic steps of an image obtained using a scanning probe microscope.
We adopt a mixture model consisting of multiple statistical distributions to describe an image.
This statistical approach provides a comprehensive way to subtract a background surface even in the presence of atomic steps as well as to evaluate terrace heights in a single framework.
Moreover, it also enables us to extract further quantitative information by introducing additional prior knowledge about the image.
An example of this extension is estimating a unit height of atomic steps together with the terrace heights.
We demonstrate the capability of our method for a topographic image of a Cu(111) surface taken using a scanning tunneling microscope.
The background subtraction corrects all terraces to be parallel to a horizontal plane and the precision of the estimated unit height reaches the order of a picometer.
An open-source implementation of our method is available on the web.
\end{abstract}

\maketitle

\section{Introduction}

Scanning probe microscopes are ubiquitously used to measure and analyze various properties~\cite{Wiesendanger,Meyer}.
The probe is driven typically by piezo actuators that are deformed by applying voltages. This actuation mechanism realizes high spatial resolution down to atomic resolution. At the same time, however, observed images are often distorted due to non-linear response, hysteresis, and creep of piezo actuators as well as thermal drift and fluctuations in the environment around a scanning head.
Such distortions are obstacles to subsequent analysis and must be corrected.

The image correction in the surface-normal direction, or background subtraction, is usually performed using a least-squares fit.
It is simple and well-defined in the absence of atomic steps.
In the presence of atomic steps, however, even such an elementary subtraction can not be done in a straightforward manner.
One needs to classify image pixels as belonging to one of the terraces in advance of the background subtraction.
Even if one can do so, a subtracted result is subject to the rather arbitrary classification of pixels.
Nevertheless, the background subtraction in the presence of atomic steps is of quantitative importance to evaluate the heights of atomic steps, which not only carry a crucial information to identify surface terminations but also are exploited to calibrate piezo actuators in the surface-normal direction.

Here we present a statistical method for the background subtraction working regardless of the presence or absence of atomic steps.
The key idea is a mixture model consisting of a linear combination of multiple statistical distributions to describe an image.
By incorporating a background surface into the model as parameters of the distributions, the labeling of terraces, subtraction of a background surface, and estimation of terrace heights are all simultaneously achieved as a result of maximum likelihood estimation.
Moreover, the statistical approach enables us to estimate another quantity by introducing a prior information.
We estimate a unit height of atomic steps as an exemplary extension.
The unit height is an inter-plane distance between adjacent planes normal to the surface and is a dimension necessary for the calibration of the piezo actuator in the surface-normal direction.
We apply our method to a topographic image of a Cu(111) surface taken with a scanning tunneling microscope.
We demonstrate that the background is well subtracted and that the unit height can be estimated with high precision at the picometer order.
Our code, implemented not only for the demonstration but also for generic images and written in Python, is available as an open-source software~\cite{source}.

\section{Methods}

\subsection{Model}

\begin{figure}[t]
	\centering
	\includegraphics{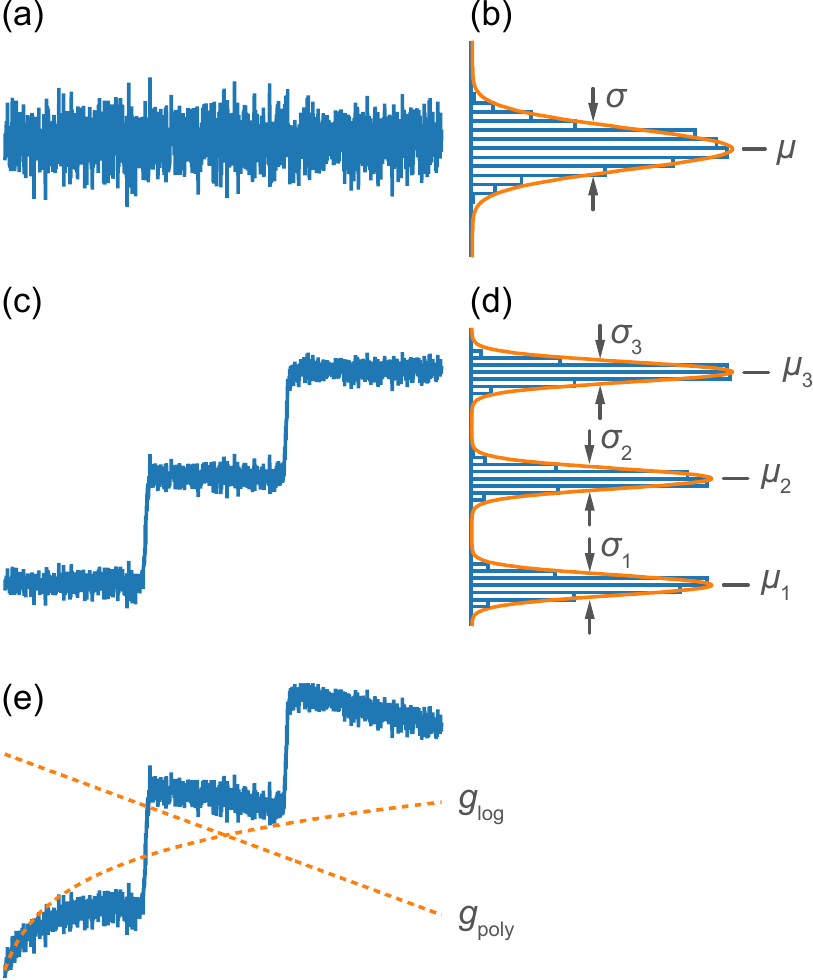}
	\caption{A schematic introduction of our model.
	The blue curves in (a), (c), and (e) represent topographic images.
	(a) No step, no tilt, and no distortion.
	(c) Steps are included in (a).
	(e) A tilt and a distortion are included in (c).
	The orange broken lines depict the tilt and the distortion.
	(b), (d) Histograms of (a) and (c), respectively.
	The orange curves indicate the distributions $f$ in Eq.~\eqref{eqn:likelihood_simplest} and $\sum_m\pi_mf$ in Eq.~\eqref{eqn:likelihood_steps}, respectively.
	}
	\label{fig:schematic}
\end{figure}

To formulate our model, we begin with a simple situation: no step, no tilt, and no distortion (Fig.~\ref{fig:schematic}(a)).
Corrugations in an image then solely stem from atoms, impurities, and random noise.
The heights distribute around a central value and a histogram of the heights can be approximated by the probability density function of a statistical distribution $f$ (Fig.~\ref{fig:schematic}(b)).
As a basic assumption, we regard that height at each pixel is independently generated from $f$.
Let $t_n$ be height at $n$-th pixel of an image, where $n$ is numbered from 1 to $N$ in the chronological order of a measurement.
$N$ is the total number of pixels in the image.
The likelihood function to obtain the image is
\begin{align}
	p(\bm{t}|\mu, \sigma) = \prod_{n=1}^Nf(t_n|\mu, \sigma),
	\label{eqn:likelihood_simplest}
\end{align}
where $\bm{t}=(t_1,\cdots, t_N)^\top$ is a vector representation of the image, and $\mu$ and $\sigma$ are the location and scaling parameters of $f$, respectively.
Although $f$ can generally have more parameters, we consider the normal distribution and the Cauchy distribution in this paper for simplicity, written as
\begin{equation}
	f(t_n|\mu, \sigma) =
	\begin{cases}
		\dfrac{1}{\left(2\pi\sigma^2\right)^{1/2}}\exp\left\{-\dfrac{(t_n-\mu)^2}{2\sigma^2}\right\} & \mathrm{(normal)}\\
		\dfrac{1}{\pi}\dfrac{\sigma}{(t_n-\mu)^2+\sigma^2} & \mathrm{(Cauchy)}
	\end{cases}
\end{equation}
$\mu$ and $\sigma$ mean height of the terrace and magnitude of corrugations of the terrace, respectively.
The normal distribution is considered because of its relevance to least-squares fit and the Cauchy distribution is chosen as a heavy-tail distribution, the importance of which we will see later.

Next we consider steps and terraces.
Suppose that there are $M$ terraces different in height in the image (Fig.~\ref{fig:schematic}(c) and Fig.~\ref{fig:schematic}(d)).
Steps and terraces are introduced by extending Eq.~\eqref{eqn:likelihood_simplest} to a linear combination of $f$,
\begin{equation}
	p(\bm{t}|\bm{\mu}, \bm{\sigma}, \bm{\pi}) = \prod_{n=1}^N
	\sum_{m=1}^M\pi_m f\left(t_n\left|\mu_m, \sigma_m\right.\right),
	\label{eqn:likelihood_steps}
\end{equation}
where $\bm{\mu}=(\mu_1\ \mu_2\ \cdots\ \mu_M)^\top$ and $\bm{\sigma}=(\sigma_1\ \sigma_2\ \cdots\ \sigma_M)^\top$ are the location and scale parameters of terraces, respectively.
$\bm{\pi}=(\pi_1\ \pi_2\ \cdots\ \pi_M)^\top$ are mixing coefficients satisfying constraints
\begin{equation}
	\pi_m > 0, \quad \sum_{m=1}^M\pi_m=1.
	\label{eqn:constraints}
\end{equation}

Now we include a tilt and distortions.
We consider positional and chronological terms (Fig.~\ref{fig:schematic}(e)).
As positional terms, we adopt a polynomial surface, $g_\mathrm{poly}(\bm{w}; \bm{r}_n) = \bm{w}^\top\bm{\phi}(\bm{r}_n)$, where $\bm{w}=(w_0, w_1, w_2, \cdots)^\top$ and $\bm{\phi}(x, y)=(1, x, y, x^2, xy, y^2, \cdots)^\top$ are polynomial coefficients and bases, respectively, and $\bm{r}_n=(x_n, y_n)^\top$ is the lateral position of the $n$-th pixel.
The linear term denotes a tilt and the higher-order terms represents non-linear distortions.
Chronological terms describe creep of a piezo actuator diminishing logarithmically in time~\cite{Meyer}, $g_\mathrm{log}(\bm{A}, \bm{\tau}; n)=\sum_{j=1}^J A_j\ln(n+\tau_j)$, where $\bm{A}=(A_1\ A_2\ \cdots\ A_J)^\top$ and $\bm{\tau}=(\tau_1\ \tau_2\ \cdots\ \tau_J)^\top$ are scale and decay parameters, respectively.
These terms are included in Eq.~\eqref{eqn:likelihood_steps} as
\begin{align}
	p(\bm{t}|\bm{\theta}) &= \prod_{n=1}^N \sum_{m=1}^M\pi_mf_{mn},
	\label{eqn:likelihood_full}\\
	f_{mn} &=
		f\left(t_n\left|\mu_m+g_\mathrm{poly}(\bm{w}; \bm{r}_n)+g_\mathrm{log}(\bm{A}, \bm{\tau}; n), \sigma_m\right.\right),
	\label{eqn:likelihood_full_details}
\end{align}
where $\bm{\theta}$ is a set of parameters, $\bm{\theta}=\{\bm{\mu}, \bm{\sigma}, \bm{w}, \bm{A}, \bm{\tau}, \bm{\pi}\}$, introduced for brevity of notation.
The degree that $n$-th pixel belongs in terrace $m$ is given by the responsibility~\cite{PRML},
\begin{equation}
	\gamma_{mn} = \dfrac{\pi_mf_{mn}}{\sum_{m^\prime=1}^M\pi_{m^\prime}f_{m^\prime n}}.
	\label{eqn:responsibility}
\end{equation}
The responsibility not only works to label terraces but also plays an important role for calculation of the maximum likelihood estimation as described below.

\subsection{Background subtraction and unit height of atomic steps}

Background subtraction, or removal of the tilt and distortions, is achieved by
\begin{equation}
	t_n - g_\mathrm{poly}(\hat{\bm{w}};\bm{r}_n)-g_\mathrm{log}(\hat{\bm{A}}, \hat{\bm{\tau}}; n),
	\label{eqn:background_subtraction}
\end{equation}
where the parameters with the hat ($\hat{\bm{w}}$, $\hat{\bm{A}}$, and $\hat{\bm{\tau}}$) are those maximizing the likelihood function Eq.~\eqref{eqn:likelihood_full}, \textit{i.e.}, the maximum likelihood estimate,
\begin{equation}
	\hat{\bm{\theta}} \equiv \{\hat{\bm{\mu}}, \hat{\bm{\sigma}}, \hat{\bm{w}}, \hat{\bm{A}}, \hat{\bm{\tau}}, \hat{\bm{\pi}}\} = \argmax_{\bm{\theta}} p(\bm{t}|\bm{\theta}).
	\label{eqn:solution_ML}
\end{equation}
Since all the parameters are simultaneously optimized, we can estimate terrace heights and label terraces together with the background subtraction.
When $M=1$ and $f$ is a normal distribution, this procedure is equivalent to a standard least-squares fit~\cite{PRML}.
In this sense, our model includes least-squares fit as a special case.

To estimate a unit height of atomic steps, we introduce a prior probability for $\mu_m$.
Since $\mu_m$ is a height of a terrace, it is expected to be close to an integer multiple of a unit height $c_0$ with an appropriate offset $\mu_0$,
\begin{equation}
	\mu_m \sim j_mc_0+\mu_0,
\end{equation}
where $j_m$ is an integer.
This is equivalent to
\begin{equation}
	\cos\frac{2\pi(\mu_m-\mu_0)}{c_0} \sim 1.
\end{equation}
Therefore, a prior probability for $\mu_m$ suited to estimate $c_0$ is the von Mises distribution,
\begin{equation}
	p(\mu_m|c_0, \varphi_0, \kappa) =
	\frac{1}{2\pi I_0(\kappa)}
	\exp\left\{\kappa\cos\left(\frac{2\pi\mu_m}{c_0}-\varphi_0\right)\right\},
	\label{eqn:vonMises}
\end{equation}
where $\varphi_0$ and $\kappa$ are the location parameter and the concentration parameter, respectively, and $I_0$ the modified Bessel function of order 0.
$\kappa$ is a hyperparameter controlling strength of periodicity of $\mu_m$.
An estimate of $c_0$ is obtained by maximizing the posterior probability,
\begin{equation}
	\left\{\hat{\bm{\theta}}, \hat{c}_0, \hat{\varphi}_0\right\}
	= \argmax_{\bm{\theta}, c_0, \varphi_0}
	p(\bm{t}|\bm{\theta}) \prod_{m=1}^M p(\mu_m|c_0, \varphi_0, \kappa).
	\label{eqn:solution_MAP}
\end{equation}

\subsection{Optimization}
\label{sec:optimization}
The technical goals for the background subtraction and the unit height estimation are to find a maximum likelihood solution (Eq.~\eqref{eqn:solution_ML}) and a maximum posterior solution (Eq.~\eqref{eqn:solution_MAP}), respectively.
The expectation-maximization (EM) algorithm, which is an iterative method repeating an expectation (E) step and a maximization (M) step until convergence, is a powerful method for this purpose~\cite{PRML,EM}.
We outline the procedure of optimization using the EM algorithm specifically as follows.
\begin{enumerate}
	\item Choose initial parameters.
		This is critical to prevent the iteration loop below from being caught by a meaningless local maximum.
		As it is not trivial to directly determine a set of parameters $\left\{\bm{w}^{(1)}, \bm{\mu}^{(1)}, \bm{\sigma}^{(1)}, \bm{\pi}^{(1)}\right\}$, instead define an initial responsibility $\gamma_{mn}^{(0)}$ by setting $\gamma_{mn}^{(0)}=1$ if $n$-th pixel is likely to belong in terrace $m$.
		Details are provided in Appendix \ref{sec:initial_parameters}.
	\item Calculate the objective function, which is log likelihood $\ln p(\bm{t}|\bm{\theta})$ for the background subtraction and log posterior $\ln p(\bm{t}|\bm{\theta})p(\bm{\mu}|c_0, \varphi_0, \kappa)$ for the unit height estimation.
	\item\label{item:E-step} E step.
		Calculate the responsibility $\gamma_{mn}$ (Eq.~\eqref{eqn:responsibility}) with the current set of parameters $\bm{\theta}^{(i)}$, where $i$ is the iteration index starting from 1.
	\item M step.
		For the background subtraction, find a new set of parameters given by
		\begin{align}
			\bm{\theta}^{(i+1)} &= \argmax_{\bm{\theta}} L(\bm{\theta}, \gamma_{mn}),
			\label{eqn:solution_ML_mstep}\\
			L(\bm{\theta}, \gamma_{mn}) &=
			\sum_{m=1}^M\sum_{n=1}^N\gamma_{mn}\left\{\ln\pi_m+\ln f_{mn}\right\}
			+\lambda\left(\sum_{m=1}^M\pi_m - 1\right),
		\end{align}
		where $\lambda$ is a Lagrange multiplier to take account of the constraint Eq.~\eqref{eqn:constraints}.
		For the unit height estimation, new parameters are
		\begin{equation}
			\left\{\bm{\theta}^{(i+1)}, c_0^{(i+1)}, \varphi_0^{(i+1)}\right\} =
			\argmax_{\bm{\theta}, c_0, \varphi_0} \left\{
				L(\bm{\theta}, \gamma_{mn})
				+N\sum_{m=1}^M \ln p(\mu_m|c_0, \varphi_0, \kappa)
			\right\}.
		\end{equation}
		A solution is obtained using a gradient ascent method.
		We adopt ADAM~\cite{ADAM}.
		For the background subtraction, a solution of Eq.~\eqref{eqn:solution_ML_mstep} is analytically obtained when $g_\mathrm{log}=0$.
		This solution is practically important as an initial value for general cases because it can be obtained quickly.
		See Appendix \ref{sec:analytical_solution} for details.
	\item Recalculate the objective function.
		If the value does not satisfy a convergence criteria, return to (\ref{item:E-step}).
\end{enumerate}

\subsection{Experiments}
The Cu(111) surface for the demonstration below was prepared in a ultra-high vacuum chamber by repeating Ar sputtering and subsequent annealing at 770~K.
We measure it with a scanning tunneling microscope at 4.3~K.
The tip was electrochemically etched tungsten wire cleaned by electron beam heating and a field ion microscope.
All heights in the following are nominal values with a rough calibration using the conventional method of line profiles.

\section{Results and discussion}

\begin{figure}[t]
	\centering
	\includegraphics{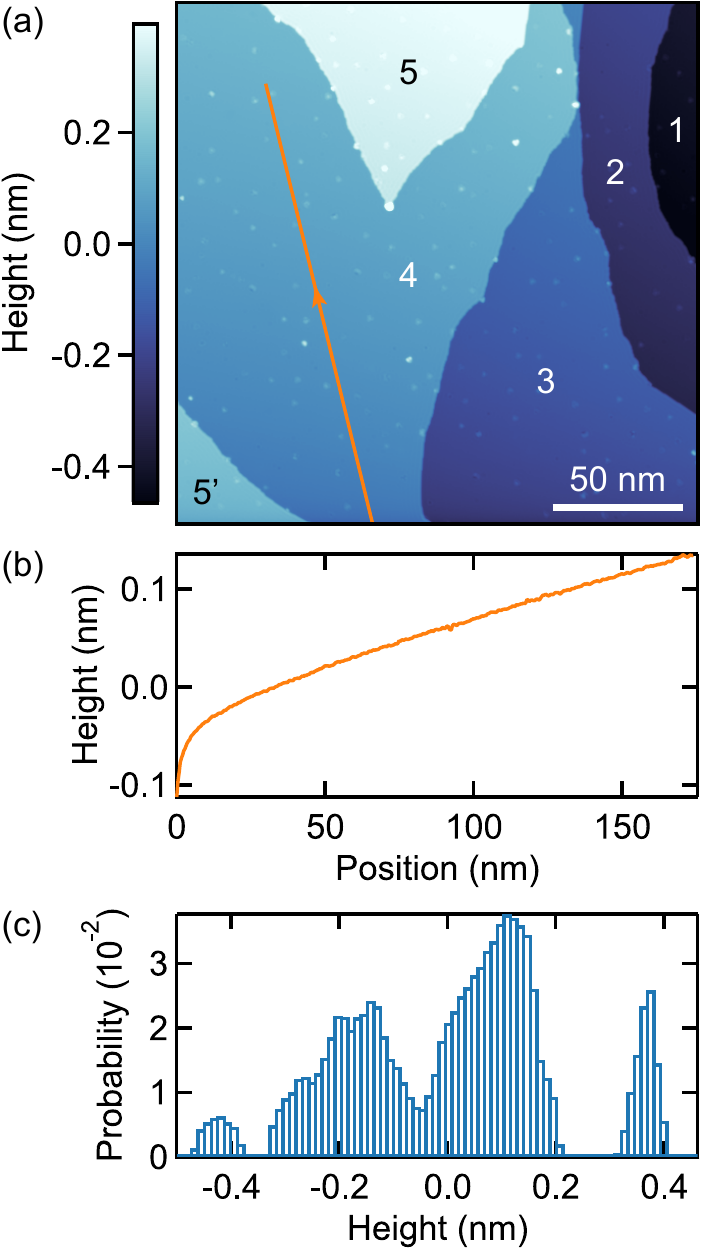}
	\caption{(a) A 200$^2$~nm$^2$ constant-current image of Cu(111).
	The sample bias voltage is -0.1~V and the tunneling current is 10~pA.
	The numbers denote terraces.
	The line shows a trajectory of a line profile of (b) and the arrow indicates the direction in which the line profile is sampled.
	(b) A line profile taken along the line shown in (a).
	(c) A histogram of the image shown in (a).}
	\label{fig:raw}
\end{figure}

To test and demonstrate the capability of our method, we apply it to an image measured with a scanning tunneling microscope.
Figure~\ref{fig:raw}(a) shows a constant-current image of a Cu(111) surface taken in a field of view of 200$^2$~nm$^2$.
The terraces 5 and 5$^\prime$ are supposed to be the same height but shown in different colors due to a tilt of the sample.
Creep of the piezo actuator in the direction normal to the surface is prominent around the bottom of the image where the scan started.
Both the tilt and the creep are highlighted in the line profile shown in Fig.~\ref{fig:raw}(b).
Because of the tilt and the distortion, the histogram of the image (Fig.~\ref{fig:raw}(c)) shows four broad peaks even though five terraces different in height exist in the image.

\begin{figure}[t]
	\centering
	\includegraphics{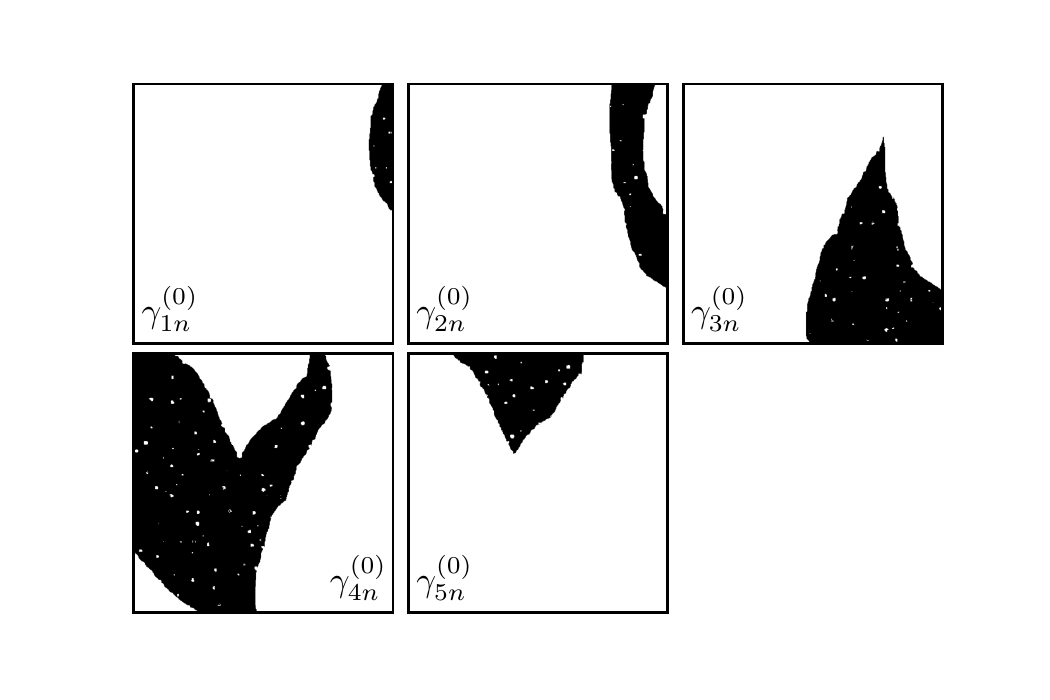}
	\caption{The initial responsibility $\gamma_{mn}^{(0)}$ obtained by clustering.
	Pixels set to 1 are shown in black.
	The clustering threshold is 10~pm.}
	\label{fig:initial_responsibility}
\end{figure}

The initial responsibility $\gamma_{mn}^{(0)}$ (Sec. \ref{sec:optimization}) is set by clustering pixels such that each cluster consists of pixels in a terrace.
This is done by clustering neighboring pixels if their difference in height is less than a threshold (Fig.~\ref{fig:initial_responsibility}).
$\gamma_{mn}^{(0)}$ at the pixels of the terrace 5$^\prime$ is intentionally left at 0 so that we can verify that the pixels of the terraces 5 and 5$^\prime$ are classified in the same class.

\begin{figure}[t]
	\centering
	\includegraphics{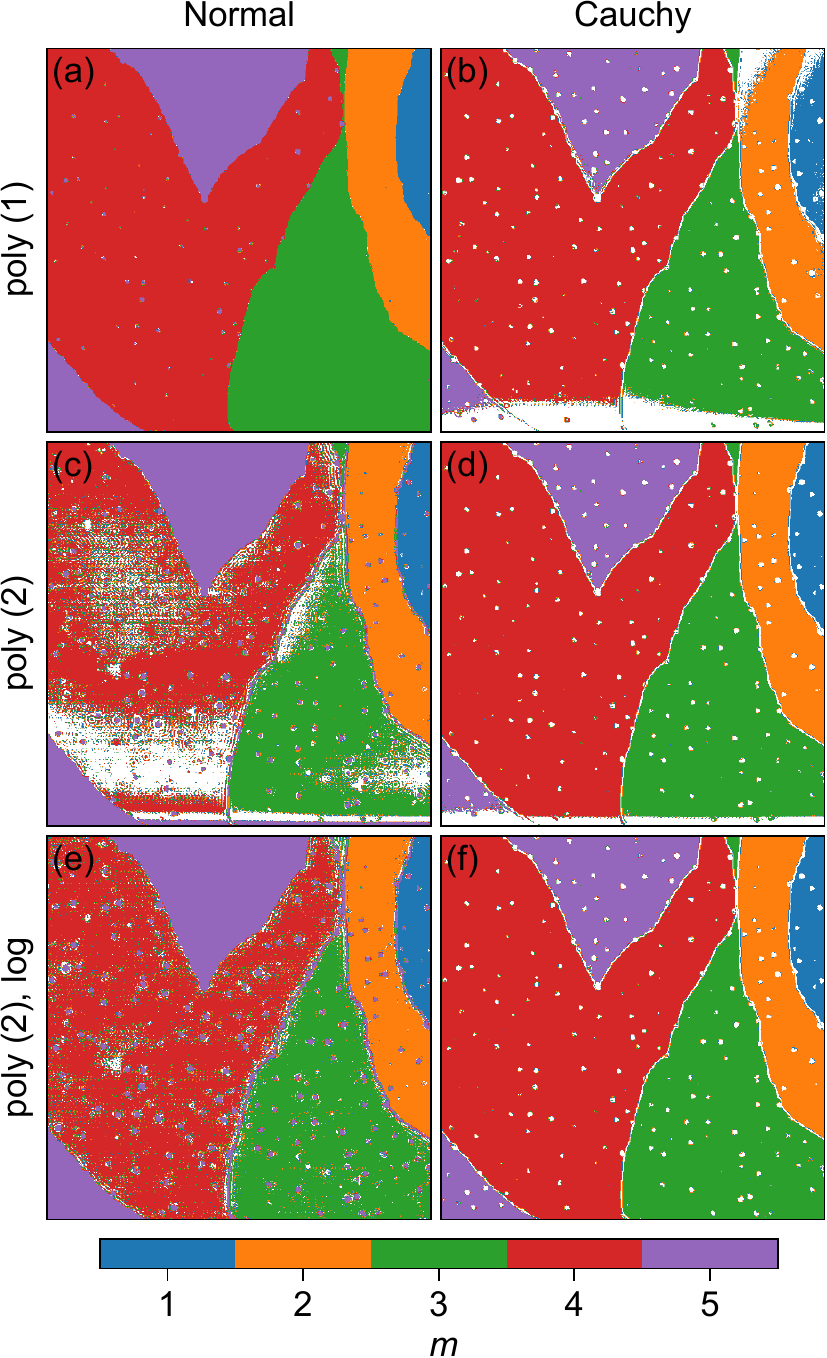}
	\caption{The optimized responsibility $\gamma_{mn}$ for different models.
	The model distribution is a normal distribution for the left column (a, c, and e) and a Cauchy distribution for the right column (b, d, and f).
	The background is a linear plane for the top row (a and b), a quadratic surface for the middle row (c and d), and a quadratic surface and log decays for the bottom row (e and f).
	The color code denotes $m$ satisfying $\gamma_{mn}>0.99$.
	Namely, $\gamma_{mn} > 0.99$ for a certain $m$ at colored pixels,
	$\gamma_{mn} \leq 0.99$ for all $m$ at white pixels. }
	\label{fig:responsibility}
\end{figure}

With the initial responsibility, we examine several different models.
The models are constructed by combinations of a normal distribution or a Cauchy distribution for the model distribution, a linear plane or a quadratic surface for the polynomial term, and no or two logarithmic decay terms~\cite{decay}.
Figure~\ref{fig:responsibility} shows the optimized responsibility $\gamma_{mn}$ (Eq.~\eqref{eqn:responsibility}) of the different models.
At colored pixels, $\gamma_{mn}$ exceeds a threshold for a certain $m$ that is depicted by the color code, whereas $\gamma_{mn}$ does not exceed the threshold for any $m$ at white pixels.
The threshold is chosen to be 0.99, meaning that a colored pixel is suggested to belong almost exclusively in a single terrace.
We note that the terraces 5 and 5$^\prime$ are shown in the same color and the other terraces are shown in different colors.
The independence of this feature to the details of the models indicates the aptitude of the statistical approach.

Meanwhile, the differences between $\gamma_{mn}$ shown in Fig.~\ref{fig:responsibility} suggest that the model distribution must be chosen carefully.
A behavior expected for an appropriate model is that pixels around impurities and steps are not classified exclusively into a terrace.
The two distributions are clearly different in this regard even for the simplest background, a linear plane and no decay as shown in Fig.~\ref{fig:responsibility}(a) and \ref{fig:responsibility}(b).
White pixels are found as expected in Fig.~\ref{fig:responsibility}(b).
Moreover, given the creep of the piezo actuator shown in Fig.~\ref{fig:raw}(b), the white pixels around the bottom of Fig.~\ref{fig:responsibility}(b) are also expected for models without the decay.
In contrast, white pixels are barely found in Fig.~\ref{fig:responsibility}(a) and impurities in the terrace 4 (red) are classified into the terrace 5 (purple).
The deviation from the requisite behaviors found in Fig.~\ref{fig:responsibility}(a) is ascribed to sensitiveness of the normal distribution to outliers.
The differences between the two distributions are enhanced by including more parameters in the model.
For the Cauchy distribution, the white pixels near the top-right corner of Fig.~\ref{fig:responsibility}(b) disappear by including a quadratic surface as shown in Fig.~\ref{fig:responsibility}(d) and those around the bottom disappear by further including logarithmic decays as shown in Fig.~\ref{fig:responsibility}(f) while pixels around impurities and steps are kept white.
For the normal distribution, however, such systematic improvements are not realized (Fig.~\ref{fig:responsibility}(c) and \ref{fig:responsibility}(e)).
These distinct behaviors suggest that a heavy-tail distribution is suited as a model distribution.
Hereafter we focus on the model of Fig.~\ref{fig:responsibility}(f).

\begin{figure}[t]
	\centering
	\includegraphics{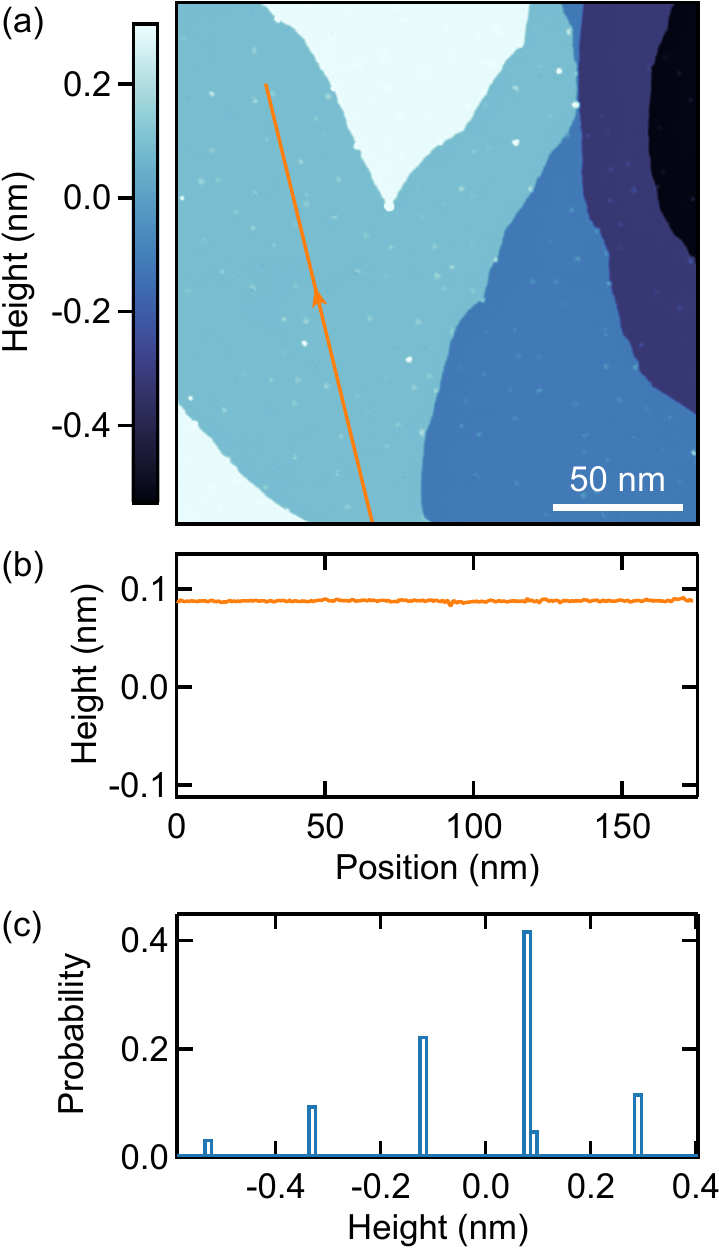}
	\caption{(a) A background-subtracted image.
	The model is the same as that of Fig.~\ref{fig:responsibility}(e).
	The raw image is the same as that of Fig.~\ref{fig:raw}(a).
	The line shows the same trajectory shown in Fig.~\ref{fig:raw}(a).
	(b) A line profile sampled along the line and in direction of the arrow shown in (a).
	The range of the vertical axis is the same as that of Fig.~\ref{fig:raw}(b).
	(c) A histogram of the image shown in (a).
	The bin width is the same as that of Fig.~\ref{fig:raw}(c).}
	\label{fig:subtracted}
\end{figure}

A background-subtracted image (Eq.~\eqref{eqn:background_subtraction}) obtained from the model of Fig.~\ref{fig:responsibility}(f) is shown in Fig.~\ref{fig:subtracted}(a).
The background is well subtracted as highlighted in the flat line profile of Fig.~\ref{fig:subtracted}(b).
The histogram of the subtracted image (Fig.~\ref{fig:subtracted}(c)) shows sharp 5 peaks at almost evenly spaced heights.
The stark contrast between the histograms of Fig.~\ref{fig:raw}(c) and Fig.~\ref{fig:subtracted}(c) signifies the importance of background subtraction for estimating the unit height of steps.

\begin{figure*}[t]
	\centering
	\includegraphics{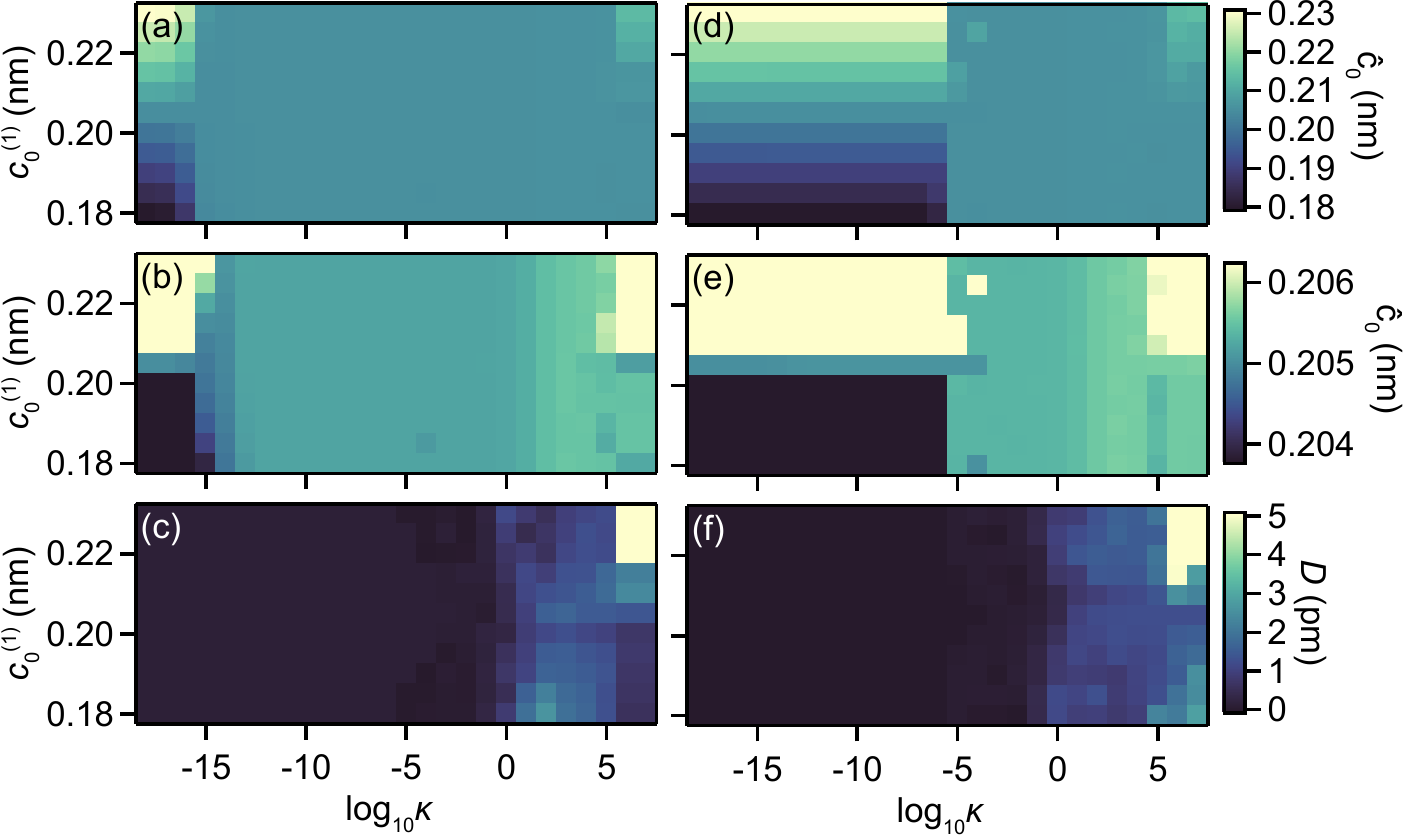}
	\caption{(a), (d) The estimated unit height of steps.
	The models are the same as used for figures~\ref{fig:responsibility}(f) and \ref{fig:responsibility}(e), respectively.
	(b), (e) The same data as (a) and (d), respectively.
	The range of color scale is adjusted to emphasize variations near the central value.
	(c), (f) The difference between $\hat{\bm{\mu}}$ and $\bm{\mu}^{(1)}$ defined in Eq.~\eqref{eqn:D}.
	The models are the same as those of (a) and (d), respectively.
	A common range of color scale is used for each row of panels.}
	\label{fig:gridsearch}
\end{figure*}

To find an appropriate range of the concentration parameter $\kappa$ for the estimation of Eq.~\eqref{eqn:solution_MAP}, we performed a grid search of the estimated unit height $\hat{c}_0$ together with the estimated terrace height $\hat{\bm{\mu}}$ for various $\kappa$ and initial value of the unit height $c_0^{(1)}$.
The result for $\hat{c}_0$ is shown in Fig.~\ref{fig:gridsearch}(a) and \ref{fig:gridsearch}(b), and for $\hat{\bm{\mu}}$ in Fig.~\ref{fig:gridsearch}(c) where we plot a measure of difference between $\hat{\bm{\mu}}$ and the initial value of $\bm{\mu}$ defined by
\begin{equation}
	D = \left(\left|\hat{\bm{\mu}} - \bm{\mu}^{(1)}\right|^2 / M\right)^{1/2}.
	\label{eqn:D}
\end{equation}
We used the maximum likelihood estimate of $\bm{\mu}$ obtained for Fig.~\ref{fig:responsibility}(f) as $\bm{\mu}^{(1)}$.
For small $\kappa$ $(\kappa < 10^{-15})$, the estimated results are almost the same as the initial values, $\hat{c}_0\sim c_0^{(1)}$ and $D\sim 0$, because little prior information is provided ($p(\bm{\mu}|c_0, \varphi_0, \kappa) \sim I_0(\kappa)$).
For large $\kappa$ $(\kappa > 10^5)$, $\hat{c}_0$ exhibits preference for $c_0^{(1)}$ probably because the optimization process stopped at a local maximum as a result of strong coupling between $c_0$ and $\bm{\mu}$.
In the wide middle range of $\kappa$ $(10^{-15} < \kappa < 10^{5})$, $\hat{c}_0$ is insensitive to both $\kappa$ and $c_0^{(1)}$.
This robustness enables us to safely conclude that the estimated unit height for this image is $\sim$0.205~nm.
The slight changes of both $\hat{c}_0$ and $D$ in the range of $10^0 < \kappa < 10^5$ reflect increase of coupling between $c_0$ and $\bm{\mu}$. 
The conclusion is unchanged even if we use the model of Fig.~\ref{fig:responsibility}(d), as shown in Fig.~\ref{fig:gridsearch}(d)--(f).
However, the parameter range where $\hat{c}_0$ is robustly estimated in Fig.~\ref{fig:gridsearch}(d) is narrower than that in Fig.~\ref{fig:gridsearch}(a).
The robustness therefore provides an indication whether the model is refined enough.

\begin{figure*}[t]
	\centering
	\includegraphics{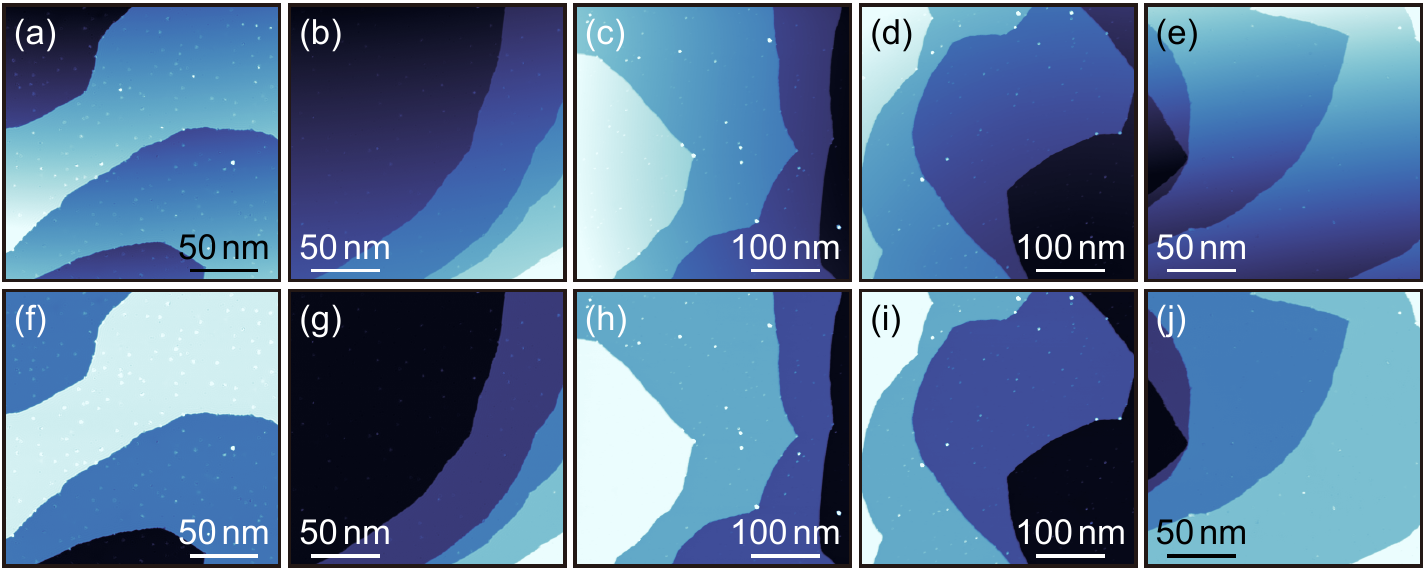}
	\caption{Images used to examine variation of the estimated unit height.
	The top panels (a)-(e) are raw images and the bottom panels (f)-(j) are corresponding subtracted images.
	The scan areas are 400$^2$~nm$^2$ for (c), (d), (h), and (i), and 200$^2$~nm$^2$ for the rest.
	The estimated unit heights are 0.204~nm for (a), 0.205~nm for (b), and 0.206~nm for (c)-(e).}
	\label{fig:images}
\end{figure*}

We note that the estimated unit height is not accurate because of the rough calibration of the piezo actuator used for the measurement but indicates the precision of our method for a given image.
To examine variation of the estimated unit height, we repeated the same analysis for different images taken using different tips but at the same temperature (Fig.~\ref{fig:images}).
We find that the unit height is estimated to be 0.205 $\pm$ 0.001~nm.
The high precision at the order of a picometer indicates that our method can take advantage of the ultra-low mechanical noise of the state-of-the-art scanning tunneling microscopes~\cite{Song10,White11,Assig13,Misra13,Roychowdhury14,Allworden18,Machida18,Balashov18,Irene18,Guan19}.

\section{Conclusion}

We have demonstrated a statistical method to remove background and to estimate a unit height of steps from a topographic image of a scanning tunneling microscope.
Our method describes an image using a mixture model.
Corrugations of a terrace are represented by a statistical distribution and multiple terraces are expressed by a linear combination of the distributions.
This statistical approach enables us to label terraces, subtract background, and estimate terrace heights.
Since nothing specific to scanning tunneling microscopes is assumed, our method can be generally applicable to topographic images taken by any other scanning probe microscopes.
Another virtue of our method is that additional prior knowledge is easily incorporated into the model.
We have estimated a unit height of atomic steps using this extension and showed that the precision of the estimated unit height reaches the order of a picometer.
This is just an example and further extensions are possible.
For example, when a thin film is grown on a substrate, film thickness can be evaluated by introducing another prior distribution.
We therefore believe that our method not only improves the subtraction method and calibration precision of scanning probe microscopes but also provides a framework to extract quantitative information.

\section*{Data availability}
The data that support the findings of this study are available from the corresponding author upon reasonable request.

\section*{Acknowledgments}
The author appreciates Christopher J. Butler for valuable suggestions and critical reading of the manuscript and also acknowledges discussions with Tetsuo Hanaguri, Tadashi Machida, and Yuuki Yasui.
This work was supported by JSPS KAKENHI Grant number 20H05280, and by MEXT as "Program for Promoting Researches on the Supercomputer Fugaku" (Basic Science for Emergence and Functionality in Quantum Matter --Innovative Strongly-Correlated Electron Science by Integration of "Fugaku" and Frontier Experiments--) (Project ID: hp200132).

\appendix
\section{\label{sec:initial_parameters}Initial parameters}

Initial parameters for the EM iteration loop are determined from the initial responsibility $\gamma_{mn}^{(0)}$ as follows.
\begin{enumerate}
	\item Calculate average height of each terrace for pixels where $\gamma_{mn}^{(0)} = 1$,
		\begin{equation}
			\mu_m^{(0)} = \frac{\sum_{n=1}^N\gamma_{mn}^{(0)}t_n}{\sum_{n=1}^N\gamma_{mn}^{(0)}}
		\end{equation}
	\item Calculate the least-squares solution of $\bm{w}$ for each terrace for pixels where $\gamma_{mn}^{(0)} = 1$ by using Eq.~\eqref{eqn:normal_w}, \eqref{eqn:normal_A}, and \eqref{eqn:normal_b},
		\begin{align}
			\bm{w}_m^{(0)} &= \left(A_m^{(0)}\right)^{-1}\bm{b}_m^{(0)},\\
			A_m^{(0)} &= \sum_{n=1}^N\gamma_{mn}^{(0)}\bm{\phi}(\bm{r}_n)\bm{\phi}(\bm{r}_n)^\top,\\
			\bm{b}_m^{(0)} &= \sum_{n=1}^N\gamma_{mn}^{(0)}\left(t_n-\mu_m^{(0)}\right)\bm{\phi}(\bm{r}_n).
		\end{align}
	\item An initial value of $\bm{w}$ is given as a weighted average of $\bm{w}_m^{(0)}$,
		\begin{equation}
			\bm{w}^{(1)} = \frac{\sum_{m=1}^M\sum_{n=1}^N\gamma_{mn}^{(0)}\bm{w}_k^{(0)}}
				{\sum_{m=1}^M\sum_{n=1}^N\gamma_{mn}^{(0)}}
		\end{equation}
	\item Initial values of $\mu_m$, $\sigma_m$, and $\pi_m$ are obtained as the least-squares solution by using Eq.~\eqref{eqn:normal_mu}, \eqref{eqn:normal_sigma}, and \eqref{eqn:normal_pi},
		\begin{align}
			\mu_m^{(1)} &= \frac{\sum_{n=1}^N
				\gamma_{mn}^{(0)}\left\{t_n-\bm{w}^{(1)\top}\bm{\phi}(\bm{r}_n)\right\}}
				{\sum_{n=1}^N\gamma_{mn}^{(0)}},\\
			\sigma_m^{(1)} &= \left[\frac{\sum_{n=1}^N
				\gamma_{mn}^{(0)}\left\{t_n-\mu_m^{(1)}-\bm{w}^{(1)\top}\bm{\phi}(\bm{r}_n)\right\}^2}
				{\sum_{n=1}^N\gamma_{mn}^{(0)}}\right]^{1/2},\\
			\pi_m^{(1)} &= \frac{\sum_{n=1}^N
				\gamma_{mn}^{(0)}}{\sum_{m=1}^M\sum_{n=1}^N\gamma_{mn}^{(0)}},
		\end{align}
\end{enumerate}

\section{\label{sec:analytical_solution}Analytical solution}

Since we consider either a normal distribution or a Cauchy distribution as the model distribution $f$, the solution of Eq.~\eqref{eqn:solution_ML_mstep} is analytically obtained by solving $\nabla_{\bm{\theta}}L=0$ if the chronological distortion $g_\mathrm{log}=0$.
When $f$ is a normal distribution, the solution is
\begin{align}
	\mu_m &= \frac{\sum_{n=1}^N\gamma_{mn}\left\{t_n-\bm{w}^\top\bm{\phi}(\bm{r}_n)\right\}}
		{\sum_{n=1}^N\gamma_{mn}},
	\label{eqn:normal_mu}\\
	\sigma_m &= \left[\frac{\sum_{n=1}^N\gamma_{mn}\left\{t_n-\mu_m-\bm{w}^\top\bm{\phi}(\bm{r}_n)\right\}^2}
		{\sum_{n=1}^N\gamma_{mn}}\right]^{1/2},
	\label{eqn:normal_sigma}\\
	\pi_m &= \frac{\sum_{n=1}^N\gamma_{mn}}{\sum_{m=1}^M\sum_{n=1}^N\gamma_{mn}},
	\label{eqn:normal_pi}\\
	\bm{w} &= A^{-1}\bm{b},\label{eqn:normal_w}\\
	A &= \sum_{m=1}^M\sum_{n=1}^N\gamma_{mn}\bm{\phi}(\bm{r}_n)\bm{\phi}(\bm{r}_n)^\top,
	\label{eqn:normal_A}\\
	\bm{b} &= \sum_{m=1}^M\sum_{n=1}^N\gamma_{mn}(t_n-\mu_m)\bm{\phi}(\bm{r}_n)
	\label{eqn:normal_b}.
\end{align}
Note that $\mu_m$ and $\bm{w}$ are mutually included in their solutions.
Each element of $\bm{\theta}^{(i+1)}$ is provided as
\begin{align}
	\mu_m^{(i+1)} &= \frac{\sum_{n=1}^N\gamma_{mn}\left\{t_n-\bm{w}^{(i)\top}\bm{\phi}(\bm{r}_n)\right\}}
		{\sum_{n=1}^N\gamma_{mn}},\\
	\sigma_m^{(i+1)} &= \left[\frac{\sum_{n=1}^N\gamma_{mn}\left\{t_n-\mu_m^{(i+1)}-\bm{w}^{(i)\top}\bm{\phi}(\bm{r}_n)\right\}^2}
		{\sum_{n=1}^N\gamma_{mn}}\right]^{1/2},\\
	\bm{w}^{(i+1)} &= A^{-1}\bm{b}^{(i+1)},\\
	\bm{b}^{(i+1)} &= \sum_{m=1}^M\sum_{n=1}^N\gamma_{mn}\left(t_n-\mu_m^{(i+1)}\right)\bm{\phi}(\bm{r}_n).
\end{align}
When $f$ is a Cauchy distribution, the solution is,
\begin{align}
	\mu_m &= \frac{\sum_{n=1}^N\tilde{\gamma}_{mn}\left\{t_n-\bm{w}^\top\bm{\phi}(\bm{r}_n)\right\}}
		{\sum_{n=1}^N\tilde{\gamma}_{mn}},\\
	\sigma_m &= \frac{\sum_{n=1}^N\gamma_{mn}}{2\pi\sum_{n=1}^N\tilde{\gamma}_{mn}},\\
	A &= \sum_{m=1}^M\sum_{n=1}^N\frac{\tilde{\gamma}_{mn}}{\sigma_m}
		\bm{\phi}(\bm{r}_n)\bm{\phi}(\bm{r}_n)^\top,\\
	\bm{b} &= \sum_{m=1}^M\sum_{n=1}^N\frac{\tilde{\gamma}_{mn}}{\sigma_m}
		(t_n-\mu_m)\bm{\phi}(\bm{r}_n),\\
	\tilde{\gamma}_{mn} &= \gamma_{mn}f\!\left(t_n\!\left|\mu_m+\bm{w}^\top\bm{\phi}(\bm{r}_n\right.\right),\sigma_m).
\end{align}
$\pi_m$ and $\bm{w}$ are given as the same formulae as those of a normal distribution, \eqref{eqn:normal_pi} and \eqref{eqn:normal_w}, respectively.
Each element of $\bm{\theta}^{(i+1)}$ is provided as
\begin{align}
	\tilde{\gamma}_{mn}^{(i+1)} &= \gamma_{mn}f\!\left(t_n\!\left|\mu_m^{(i)}+\bm{w}^{(i)\top}\bm{\phi}(\bm{r}_n),\sigma_m^{(i)}\right.\right)\\
	\mu_m^{(i+1)} &= \frac{\sum_{n=1}^N\tilde{\gamma}_{mn}^{(i+1)}\left\{t_n-\bm{w}^{(i)\top}\bm{\phi}(\bm{r}_n)\right\}}
		{\sum_{n=1}^N\tilde{\gamma}_{mn}^{(i+1)}},\\
	\sigma_m^{(i+1)} &= \frac{\sum_{n=1}^N\gamma_{mn}}{2\pi\sum_{n=1}^N\tilde{\gamma}_{mn}^{(i+1)}},\\
	\bm{w}^{(i+1)} &= \left(A^{(i+1)}\right)^{-1}\bm{b}^{(i+1)},\\
	A^{(i+1)} &= \sum_{m=1}^M\sum_{n=1}^N\frac{\tilde{\gamma}_{mn}^{(i+1)}}{\sigma_m^{(i+1)}}
		\bm{\phi}(\bm{r}_n)\bm{\phi}(\bm{r}_n)^\top,\\
	\bm{b}^{(i+1)} &= \sum_{m=1}^M\sum_{n=1}^N\frac{\tilde{\gamma}_{mn}^{(i+1)}}{\sigma_m^{(i+1)}}
		\left(t_n-\mu_m^{(i+1)}\right)\bm{\phi}(\bm{r}_n).
\end{align}

\bibliography{main}

\begin{thebibliography}{17}
\bibitem [{dec()}]{decay}%
  \BibitemOpen
  \href@noop {} {}\bibinfo {note} {{Two decay terms are considered because the
  piezo actuator of the scanning tunneling microscope used to take the image is
  split into two sectors for motion in the surface-normal
  direction.}}\BibitemShut {Stop}%
\end{thebibliography}

\end{document}